\documentclass[prl,twocolumn,showpacs,preprintnumbers,amsmath,amssymb,superscriptaddress]{revtex4-2}

\usepackage{graphicx}
\usepackage{bm}
\usepackage{bbm}
\usepackage{dsfont}
\usepackage{hyperref}
\usepackage{dcolumn}
\usepackage[mathlines]{lineno}
\usepackage{color}

\begin{document}

\title{Universal time-dependent control scheme for realizing arbitrary linear bosonic transformations}

\author{Ze-Liang \surname{Xiang}}
\affiliation{School of Physics, Sun Yat-sen University, Guangzhou 510275, China.}
\author{Diego Gonz\'{a}lez \surname{Olivares}}
\author{Juan Jos\'{e} \surname{Garc\'{i}a-Ripoll}}
\affiliation{Instituto de F\'{i}sica Fundamental IFF-PCSIC, Calle Serrano 113b, E-28006 Madrid, Spain.}
\author{Peter \surname{Rabl}}
\affiliation{Vienna Center for Quantum Science and Technology, Atominstitut, TU Wien, Stadionallee 2, 1020 Vienna, Austria.}

\begin{abstract}
We study the implementation of arbitrary excitation-conserving linear transformations between two sets of $N$ stationary bosonic modes, which are connected through a photonic quantum channel. By controlling the individual couplings between the modes and the channel, an initial $N$-partite quantum state in register $A$ can be released as a multiphoton wave packet and, successively, be reabsorbed in register $B$. Here we prove that there exists a set of control pulses that implement this transfer with arbitrarily high fidelity and, simultaneously, realize a prespecified $N\times N$ unitary transformation between the two sets of modes. Moreover, we provide a numerical algorithm for constructing these control pulses and discuss the scaling and robustness of this protocol in terms of several illustrative examples. By being purely control-based and not relying on any adaptations of the underlying hardware, the presented scheme is extremely flexible and can find widespread applications, for example, for boson-sampling experiments, multiqubit state transfer protocols or in continuous-variable quantum computing architectures.
\end{abstract}

\maketitle

Linear unitary transformations between bosonic modes play an integral part in many quantum information processing applications. For example, by sending a multimode photonic Fock state through a network of linear optical elements---thereby implementing such a unitary transformation---the output distribution of the photons is exponentially hard to predict on a classical computer~\cite{aaronson2013}, but this boson-sampling problem can be simulated efficiently in a quantum experiment~\cite{spring2013,broome2013,tillmann2013,carolan2014,spagnolo2014,wang2019,arrazola2021,brod2019}. 
When combined with single-photon sources and detectors, the same unitary transformations can be used to realize a universal quantum computer according to the Knill-Laflamme-Milburn scheme~\cite{knill2001,kok2007}. Further, by encoding quantum information in continuous-variable degrees of freedom, one can benefit from efficient bosonic error correction schemes~\cite{chuang1997,gottesman2001,mirrahimi2014,michael2016}, which is currently explored in superconducting circuits~\cite{ofek2016,hu2019,campagne-ibarcq2020} and trapped ion systems~\cite{fluhmann2019}. State transfer operations between such oscillator-encoded qubits require again the implementation of unitary transformations between distant bosonic modes.

\begin{figure}[b]
    \includegraphics[width=\columnwidth]{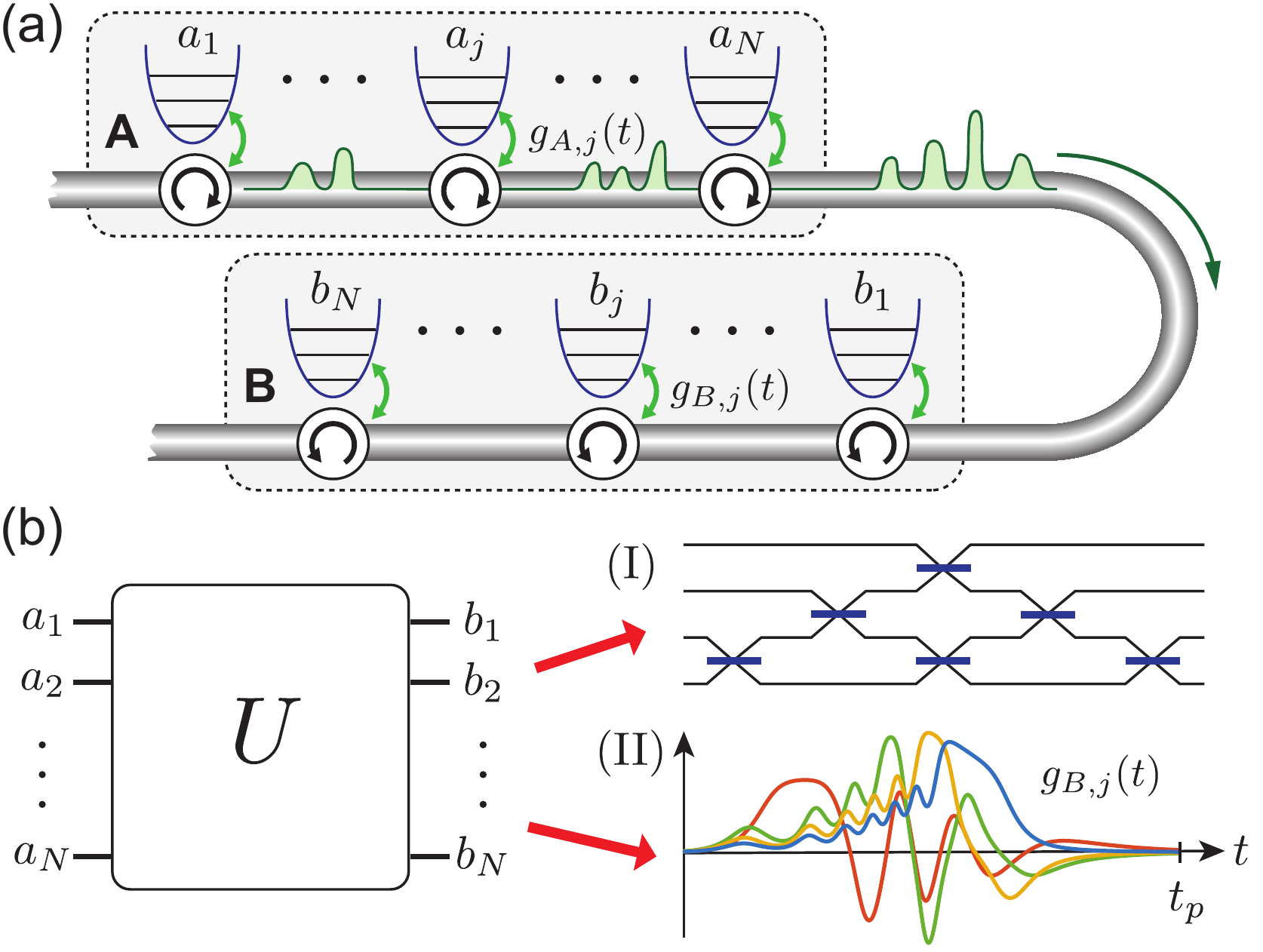}
    \caption{(a) Sketch of the quantum network considered in this Letter. Two quantum registers $A$ and $B$, each represented by  $N$ bosonic modes, are connected via a unidirectional waveguide. By controlling the couplings $g_{A,j}(t)$ and $g_{B,j}(t)$ between the modes and the waveguide, a multiphoton wave packet can be emitted from register $A$ and successively be reabsorbed in register $B$. (b) A generic linear unitary transformation $U$ between the modes, which is conventionally implemented by (I) sending photons through a network of $N(N-1)/2$ beam splitters, can be realized with our scheme in a time $t_p\sim N$ by (II) applying an appropriate set of control pulses.}
    \label{fig1:Setup}
\end{figure}

In most applications, linear unitary transformations are realized by sending photons through a network of beam splitters and phase shifters~\cite{reck1994}, with a limited amount of tunability. In this Letter, we describe a universal protocol to achieve the same task through a controlled multiphoton emission and reabsorption process. The basic idea behind this approach is summarized in Fig.~\ref{fig1:Setup}. Two quantum registers $A$ and $B$, which each contain $N$ bosonic modes, are connected by a unidirectional quantum channel. By controlling the coupling strength between the channel and each mode, a quantum state stored in register $A$ is released as a propagating multiphoton wave packet and reabsorbed in register $B$. Here we demonstrate that, for any given $N\times N$ unitary matrix $U$, there exists a set of control pulses such that (i) the reabsorption of the emitted photons can be achieved with arbitrarily high fidelity and (ii) the whole process implements the transformation
\begin{equation}\label{eq:UnitaryTransformation}
    b_{j}(t_f)= \sum_{k=1}^{N} U_{jk}  a_{k}(t_0).
\end{equation}
Here the  $a_{k}(t_0)$ are the bosonic annihilation operators for the modes of register $A$ at the initial time $t_0$ and the $b_{j}(t_f)$ are the corresponding operators for the modes of register $B$ at the final time $t_f$ of the transfer. Moreover, we provide a numerical recipe for constructing the appropriate control pulses and show that even for completely random unitaries the overall protocol time, $t_p=t_f-t_0\sim N$, only scales linearly with the number of modes. Therefore, the current approach offers an efficient and very flexible way to realize such transformations, where the targeted operation is fully specified by the shape of the control pulses and not by the network layout.


\emph{Quantum network dynamics.}---For the following analysis, we focus on the quantum network in Fig.~\ref{fig1:Setup}(a), where two sets of $N$ bosonic modes in register $A$ and register $B$, are coupled to a unidirectional waveguide. For now we assume that all modes have the same frequency $\omega_0$ and that they are coupled to the waveguide with tunable couplings $g_{A,j}(t)$ and $g_{B,j}(t)$, respectively, where $j=1,\dots, N$ labels the modes within each register. Various schemes for realizing such tunable couplings have already been demonstrated, both in the optical~\cite{keller2004,hammerer2010,nisbet-jones2011,ritter2012} and in the microwave regimes~\cite{yin2013,pierre2014,pechal2014,flurin2015,andrews2015,wulschner2016,pfaff2016,bienfait2019,dassonneville2020}.  When combined with coherent circulators~\cite{sliwa2015,kerckoff2015,chapman2017,lecocq2017,masuda2019,wang2021}, chiral waveguides~\cite{lodahl2017}, or other types of directional couplers~\cite{guimond2020,gheeraert2020,kannan2022} a fully cascaded network, as assumed in this Letter, can be implemented.

Under the assumption that the spectrum of the waveguide is sufficiently broad and approximately linear, we can adiabatically eliminate the dynamics of the propagating photons and derive a set of cascaded quantum Langevin equations for the register modes~\cite{gardiner1985,quantum_noise}. In a frame rotating with $\omega_0$, we obtain
\begin{eqnarray} \label{eq:QLE}
    \dot{c}_\mu (t)&=& -\frac{ |g_\mu(t)|^2}{2} c_\mu(t)
    - g_\mu(t)  f_{{\rm in},\mu}(t),
\end{eqnarray}
together with the input-output relations
\begin{equation}
    f_{{\rm out},\mu}(t) = f_{{\rm in},\mu}(t) + g_\mu^*(t) c_\mu (t).
\end{equation}
Here, the index $\mu$ runs over all $2N$ modes and we have made the identifications $c_\mu\equiv a_\mu$ and $g_\mu\equiv g_{A,\mu}$ for $\mu=1,\dots,N$ and $c_\mu\equiv b_{\mu-N}$ and $g_\mu\equiv g_{B,\mu-N}$ for $\mu=N+1,\dots,2N$. The in field $f_{{\rm in},1}(t)\equiv f_{\rm in}(t)$ is a $\delta$-correlated noise operator, which satisfies $[f_{{\rm in}}(t),f^\dag_{{\rm in}}(t')]=\delta(t-t')$. All other in fields are determined by the relation
$f_{{\rm in},\mu}(t)=f_{{\rm out},\mu-1}(t)$, which captures the directional nature of the quantum channel. By iterating this relation and adopting a vector notation, $\vec{c} = \left(c_1,\ldots,c_{2N}\right)^T$ and $\vec g=\left(g_1,\ldots,g_{2N}\right)^T$, we obtain
\begin{equation} \label{eq:QLEs matrix form}
    \dot{\vec{c}}(t) = - \mathcal{M} (t) \vec{c}(t) - \vec{g}(t) f_{{\rm in}}(t),
\end{equation}
where $\mathcal{M}_{\mu\nu} (t) = g_\mu(t) g^{*}_\nu(t) \Theta(\mu-\nu) $ and $\Theta(x)$ is the Heaviside function. Unless stated otherwise, we express time in units of $\gamma_{\rm max}^{-1}$, where $\gamma_{\rm max}$ denotes the maximal decay rate into the channel and depends on the specific physical implementation.  With this convention, the couplings $g_\mu(t)$ are complex numbers and constrained to $|g_\mu(t)|\leq 1$. A detailed derivation of Eq.~\eqref{eq:QLEs matrix form} can be found in the Supplemental Material~\cite{supplMultimodeU}.

The general solution of Eq.~\eqref{eq:QLEs matrix form} can be written as
\begin{equation} \label{eq:General solution}
    \vec{c}(t) = \mathcal{G}(t,t_0) \vec{c} (t_0) - \int_{t_0}^t  ds \, \mathcal{G}(t,s) \vec g(s) f_{{\rm in}}(s),
\end{equation}
where the Green's function $\mathcal{G}(t,t_0)$ obeys $\partial_t \mathcal{G}(t,t_0)=- \mathcal{M}(t) \mathcal{G}(t,t_0)$ and $\mathcal{G}(t_0,t_0)=\mathbbm{1}_{2N}$. The cascaded structure imposed by the unidirectional waveguide implies that both $\mathcal{M}$ and $\mathcal{G}$ have a lower-triangular form, i.e.,  $\mathcal{M}_{\mu\nu},\mathcal{G}_{\mu\nu}=0$ for $\mu<\nu$. Moreover, each row $\mu$ of these matrices only depends on the couplings $g_{\nu}(t)$ associated with that and previous modes $\nu\leq \mu$. This allows us to write the Green's function as
\begin{equation}
    \mathcal{G}=\left(
    \begin{array}{cc}
            G_{AA} & 0      \\
            G_{BA} & G_{BB}
        \end{array}\right) \longrightarrow \left(
    \begin{array}{cc}
            0 & 0 \\
            U & 0
        \end{array}\right),
\end{equation}
where the expression to the right indicates the targeted evolution at $t=t_f$, as specified in Eq.~\eqref{eq:UnitaryTransformation}.


\emph{Control pulses.}---To realize the desired dynamics, we first choose a set of pulse shapes for the couplings $g_{A,j}(t)$ in register $A$. These pulses do not have to be of any specific shape, but they must be mutually overlapping and satisfy~\cite{supplMultimodeU}
\begin{equation}\label{eq:LongPulses}
    \int_{t_0}^{t_f} ds \, |g_{A,j} (s)|^2\gg 1.
\end{equation}
This condition ensures that all initial excitations in register $A$ decay into the waveguide and $G_{AA}(t_f,t_0)\simeq 0$ up to exponentially small corrections.

Next, we must identify a set of control pulses $g_{B,j}(t)$, which achieve the nontrivial part of the dynamics, $G_{BA}(t_f,t_0)\rightarrow U$. To do so, we assume for now that the whole network is initially prepared in the single excitation state  $|\psi_\ell\rangle=\Psi_\ell^\dag|{\rm vac}\rangle$, where $|{\rm vac}\rangle$ is the vacuum state and $\Psi _\ell= \sum_{k=1}^N U_{\ell k} a_k(t_0)$. We then define the amplitudes $F_{j,\ell}(t,t_0)=\langle {\rm vac}| f_{{\rm out},N+j}(t)|\psi_\ell\rangle$, which represent the field in the channel right after the $j$th mode of register $B$. We obtain
\begin{equation}\label{eq:Outfields}
    F_{j,\ell}(t,t_0)= \sum_{k=1}^{N+j} g^*_{k}(t) \left[ \mathcal{G}(t,t_0) \mathcal{U}^\dag\right]_{k,\ell},
\end{equation}
where $\mathcal{U}={\rm diag}(U,0_N)$ is a block-diagonal matrix.

According to Eq.~\eqref{eq:UnitaryTransformation}, the excitation created by $\Psi^\dag_\ell$ is mapped onto the corresponding excitation of mode $b_{\ell}$ in register $B$. To achieve this mapping, during the whole protocol, the photon emitted from state $|\psi_\ell\rangle$ must not propagate beyond the $\ell$th node of register $B$, as otherwise it would be impossible to recapture it at a later time. Therefore, a necessary requirement for a perfect transfer is that the dark state condition $F_{\ell,\ell}(t,t_0)=0$ is satisfied for all times $t\in[t_0,t_f]$, or equivalently,
\begin{equation}\label{eq:DarkStateImplicit}
    g_{B,\ell}^{*}(t)\left[\mathcal{G}(t,t_0)\mathcal{U}^\dag\right]_{N+\ell,\ell}   = -F_{\ell-1,\ell}(t,t_0).
\end{equation}
For $N=1$ and $U=1$, Eq.~\eqref{eq:DarkStateImplicit} reduces to the dark-state condition employed for identifying control pulses for single-mode quantum state transfer schemes~\cite{cirac1997,jahne2007,stannigel2011,korotkov2011,xiang2017,vermersch2017,dassonneville2020} (see also Ref.~\cite{ai2021} for a preliminary extension to multimode setups).  In the Supplemental Material~\cite{supplMultimodeU}, we show that satisfying this generalized set of dark-state conditions for all $\ell=1,\dots,N$ is not only necessary, but also sufficient to obtain $G_{BA}(t_f,t_0)\simeq U$ and $G_{BB}(t_f,t_0)\simeq 0$ for sufficiently long $t_f$. Moreover, we show that the implicit equation for $g_{B,\ell}(t)$ in Eq.~\eqref{eq:DarkStateImplicit} can be converted into the following recursive expression:
\begin{equation}\label{eq:ExactSolution}
    g_{B,\ell}(t)  =
    \frac{F^*_{\ell-1,\ell}(t,t_0) }
    {\sqrt{\int_{t_0}^t ds \, | F_{\ell-1,\ell}(s,t_0)|^2}}.
\end{equation}
Because of the cascaded structure of $\mathcal{G}$, the amplitudes $F_{\ell-1,\ell}(t,t_0)$ depend on the known control pulses $g_{A,j}(t)$ and on the previously obtained  pulses $g_{B,j}(t)$ for $j<\ell$ only. Therefore, Eq.~\eqref{eq:ExactSolution} can be iteratively applied to compute all control pulses $g_{B,j}(t)$ for register $B$.

Equation~\eqref{eq:ExactSolution} proves the existence of a solution to our control problem by an explicit construction of the coupling pulses, which is the main result of this Letter. We still need to show, however, that this formal result does not lead to solutions that violate the constraints $|g_j(t)|^2 \leq 1$, are unbounded in time, or otherwise unphysical. In the following we achieve this conclusion by simply applying the protocol for engineering generic $N\times N$ unitary transformations. This approach will also allow us to deduce the scaling and the robustness of the protocol under realistic conditions.

\begin{figure}[t]
    \includegraphics[width=\columnwidth]{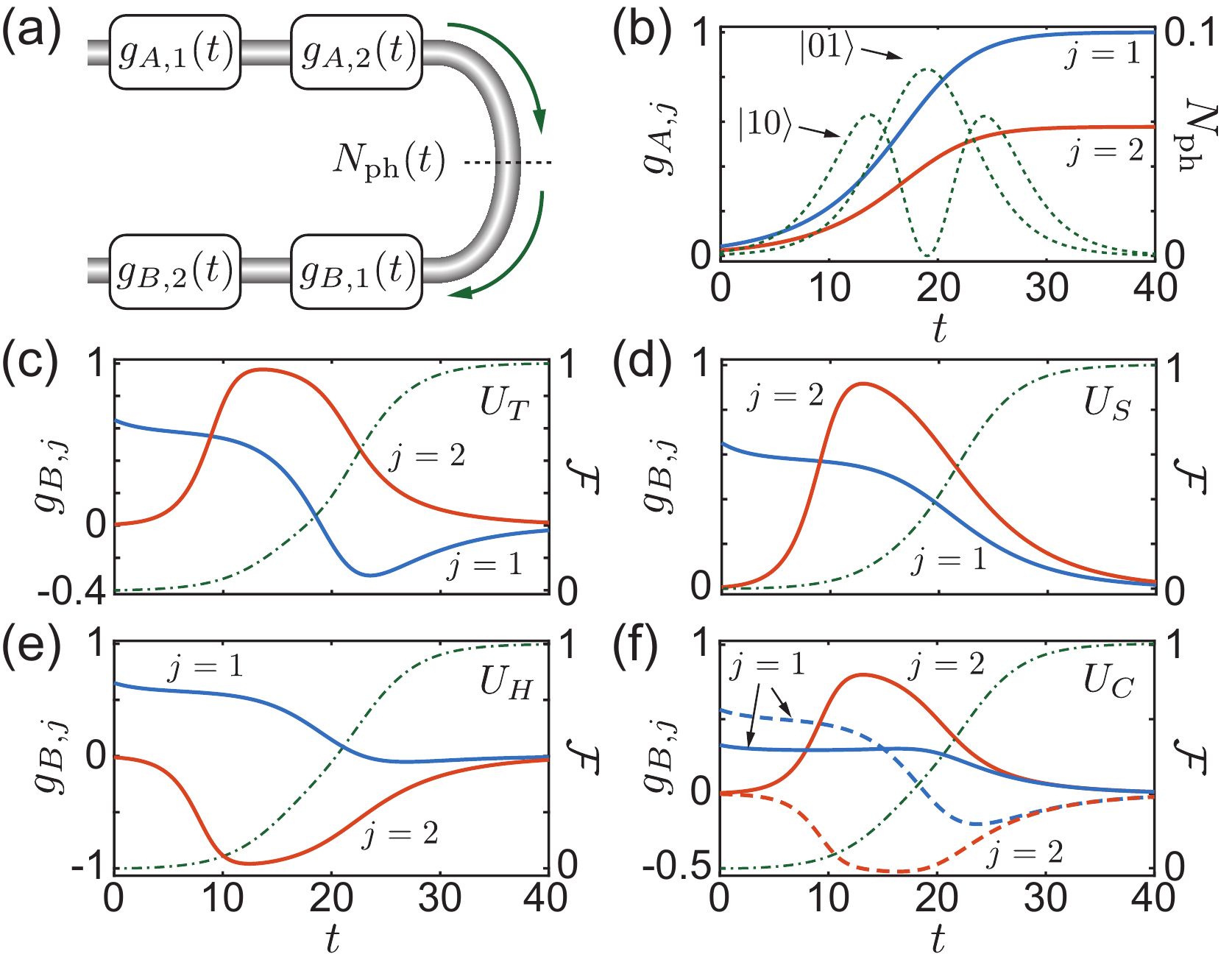}
    \caption{(a) Sketch of the setup for implementing unitary operations between $N=2$ modes. (b) Shape of the control pulses $g_{A,j}(t)$ as specified in Eq.~\eqref{eq:Pulses} and for the parameters $t_c=19$, $\delta=2$, and $\tau=1$. The dotted lines show the shape of the photon wave packet released from register $A$,  $N_{\rm ph}(t)=|\langle {\rm vac}| f_{{\rm out},2}(t)|\psi_0\rangle|^2$, for different initial states $|\psi_0\rangle=|10\rangle$ and $|\psi_0\rangle=|01\rangle$. (c)--(f) Shapes of the optimal control pulses $g_{B,j}(t)$ for the unitaries (c) $U_T$, (d) $U_S$, (e) $U_H$, and (f) $U_C$. In all plots, the dashed-dotted lines show the fidelity $\mathcal{F}$. (f) The solid lines represent the real part and the dashed lines the imaginary parts of the control pulses.}
    \label{fig2:U2x2}
\end{figure}

\emph{Two-by-two unitaries.}---In a first step, we illustrate the application of the protocol for the simplest nontrivial scenario, $N=2$, shown in Fig.~\ref{fig2:U2x2}(a). For this setup, we consider the four unitary operations
\begin{eqnarray}\label{eq:2x2Unitaries}
    U_T&=&\left(
    \begin{array}{cc}
            1 & 0 \\
            0 & 1
        \end{array}\right), \qquad
    U_S=\left(
    \begin{array}{cc}
            0 & 1 \\
            1 & 0
        \end{array}\right),\\
    U_H&=&\frac{1}{\sqrt{2}}\left(
    \begin{array}{cc}
            1 & 1  \\
            1 & -1
        \end{array}\right), \qquad U_C=\frac{1}{\sqrt{2}}\left(
    \begin{array}{cc}
            1 & i \\
            i & 1
        \end{array}\right).\nonumber
\end{eqnarray}
Here, $U_T$ corresponds to a simple state transfer between the two registers, $U_S$ additionally swaps the two modes, and the Hadamard operation $U_H$ and the unitary $U_C$ create superpositions between the modes with real and complex coefficients.

To calculate the control pulses for realizing each unitary, we set $t_0=0$ and fix the control pulses $g_{A,j}(t)$ to be of the form
\begin{equation}\label{eq:Pulses}
    g_{A,j}(t)=\frac{\eta_j}{\sqrt{e^{(t_c-t)/\tau}+1}}, \qquad \eta_j = \sqrt{\frac{1+(N-j)\delta}{1+(N-1)\delta}}.
\end{equation}
The parameters $\delta$, $t_c$, and $\tau$ can be used to optimize the protocol for a given application, but none of the following  findings depend crucially on this specific pulse shape nor on a specific set of parameters. The pulses $g_{A,j}(t)$ used in the following examples are depicted in Fig.~\ref{fig2:U2x2}(b).

Given $g_{A,j}(t)$ and $U$, we iteratively solve Eq.~\eqref{eq:ExactSolution} numerically to obtain the control pulses $g_{B,j}(t)$ and evaluate the fidelity of the operation~\cite{nielsen2002,pedersen2007}
\begin{equation}\label{eq:Fidelity}
    \mathcal{F}(t) =\frac{|{\rm Tr}\{U^\dag G_{BA}(t,t_0)\}|^2+{\rm Tr}\{G^\dag_{BA}(t,t_0)G_{BA}(t,t_0)\}}{N(N+1)} .
\end{equation}
It reaches a value of $\mathcal{F}(t_f)\simeq 1$, if the protocol was successful. For the examples in Eq.~\eqref{eq:2x2Unitaries}, the resulting pulse shapes and fidelities are plotted in Figs.~\ref{fig2:U2x2}(c)--\ref{fig2:U2x2}(f). We see that in all cases the algorithm provides the correct control pulses and the unitary transformation is implemented with close to unit fidelity, as long as the protocol time $t_p=t_f-t_0$ is long enough. We emphasize that, while the shape of the wave packet released from register $A$ depends on the initial quantum state [see Fig.~\ref{fig2:U2x2}(b)], the implemented unitary $U$ is independent of this state.


\emph{Scalability.}---Using the construct from Ref.~\cite{reck1994}, a sequential combination of $\mathcal{O}(N^2)$ of the $2\times2$ unitary operations demonstrated above is sufficient to recreate any possible $N\times N$ unitary transformation $U$ in a time $t_p\sim\mathcal{O}(N^2)$. This strategy is usually employed for implementing bosonic unitaries with photons or also atoms~\cite{chen2022,robens2022}.
However, in the current approach, already in a single run, each of the emitted photons interacts with multiple modes in register $B$. This intrinsic parallelization allows us to improve over the scheme by Reck {\it et al.}~\cite{reck1994} and obtain protocol times that only scale linearly with the number of modes, $t_p\sim\mathcal{O}(N)$.

\begin{figure}[t]
    \includegraphics[width=\columnwidth]{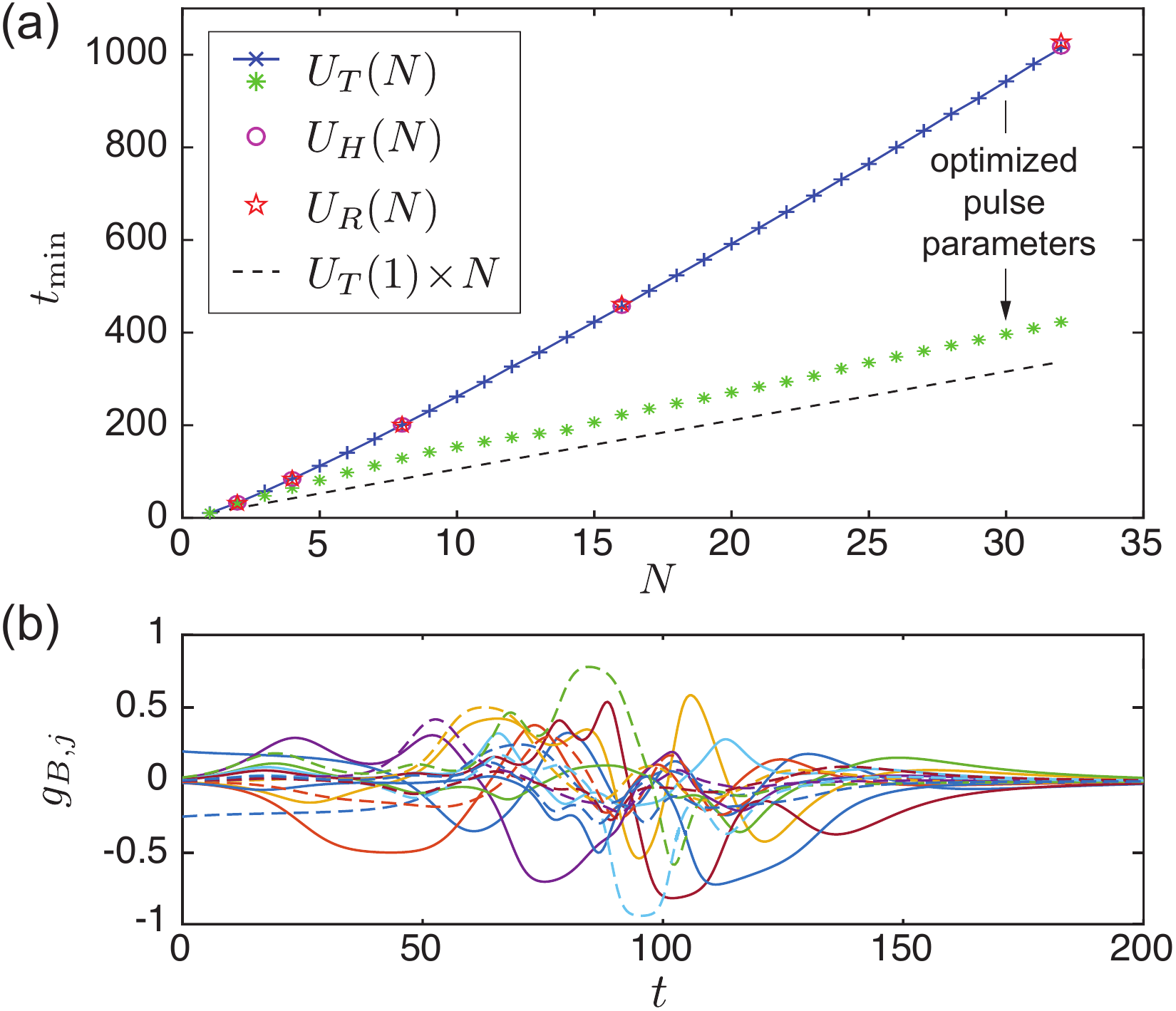}
    \caption{(a) Scaling of the minimal protocol time $t_{\rm min}$ for implementing different classes of $N\times N$ unitaries with a fidelity $\mathcal{F}\geq 0.99$. Here, $U_T(N)\equiv \mathbbm{1}_N$ is the state-transfer operation, $U_H(N)$ is the $N$-dimensional Hadamard transformation and $U_R(N)$ is a random unitary $N\times N$ matrix. In all cases, the pulses $g_{A,j}(t)$ are specified in Eq.~\eqref{eq:Pulses} with  $\delta=2$ and $\tau=1$. The results marked by stars show the minimal protocol time for $U_T(N)$ for an optimized parameter $\delta$. (b) Illustration of the numerically constructed control pulses $g_{B,j}(t)$ for the example $U_R(N=8)$. The real and imaginary parts are shown by the solid and dashed lines, respectively. See the Suppplemental Material~\cite{supplMultimodeU} for additional details.}
    \label{fig3:Scaling}
\end{figure}

To demonstrate this scaling, we numerically evaluate the minimal protocol time $t_{\rm min}$ required to implement a given $N\times N$ unitary with a fidelity of $\mathcal{F}\geq 0.99$. Specifically, we compare the implementation of the $N$-mode state-transfer operation $U_T(N)=\mathbbm{1}_{N}$, the $N$-dimensional Hadamard transformation $U_H(N)$, and generic complex unitaries $U_R(N)$ with randomly drawn matrix elements. The results are summarized in Fig.~\ref{fig3:Scaling}(a) and demonstrate that the protocol works perfectly even for a large number of modes and for arbitrary classes of unitaries. As an illustrative example,  Fig.~\ref{fig3:Scaling}(b) shows the control pulses $g_{B,j}(t)$ for $U_R(N=8)$ and qualitatively similar pulse shapes are obtained for other unitaries as well. See the Supplemental Material~\cite{supplMultimodeU} for further details about the numerical procedure that has been used to obtain these results.

The key observation from Fig.~\ref{fig3:Scaling}(a) is that $t_{\rm min}\sim N$ scales only linearly with the number of modes, independent of the specific properties of $U$. This scaling can be roughly understood as follows. Because the maximal coupling strength is bounded, $|g_j(t)|\leq \sqrt{\gamma_{\rm max}}$, the local modes can emit or absorb photons only on timescales longer than $\gamma_{\rm max}^{-1}$. Therefore, in order to emit (absorb) photons into (from) $N$ spatiotemporally distinct modes, the total pulse duration must increase proportionally to $N$. The prefactor for this scaling is the same for all the tested unitaries, but it is still a factor of $\sim 3$ higher than what one would obtain from implementing $N$ times a single-mode transfer. We attribute this overhead to the nonoptimal choice of control pulses $g_{A,j}(t)$ in Eq.~\eqref{eq:Pulses}. Indeed, for the transfer unitary $U_T(N)$, a substantial reduction of the protocol times can already be obtained by optimizing the parameter $\delta$~\cite{supplMultimodeU}. This suggests that also for other unitaries, a similar improvement of the scaling prefactor can be achieved by using other shapes for the control pulses in register $A$.

\begin{figure}[t]
    \includegraphics[width=\columnwidth]{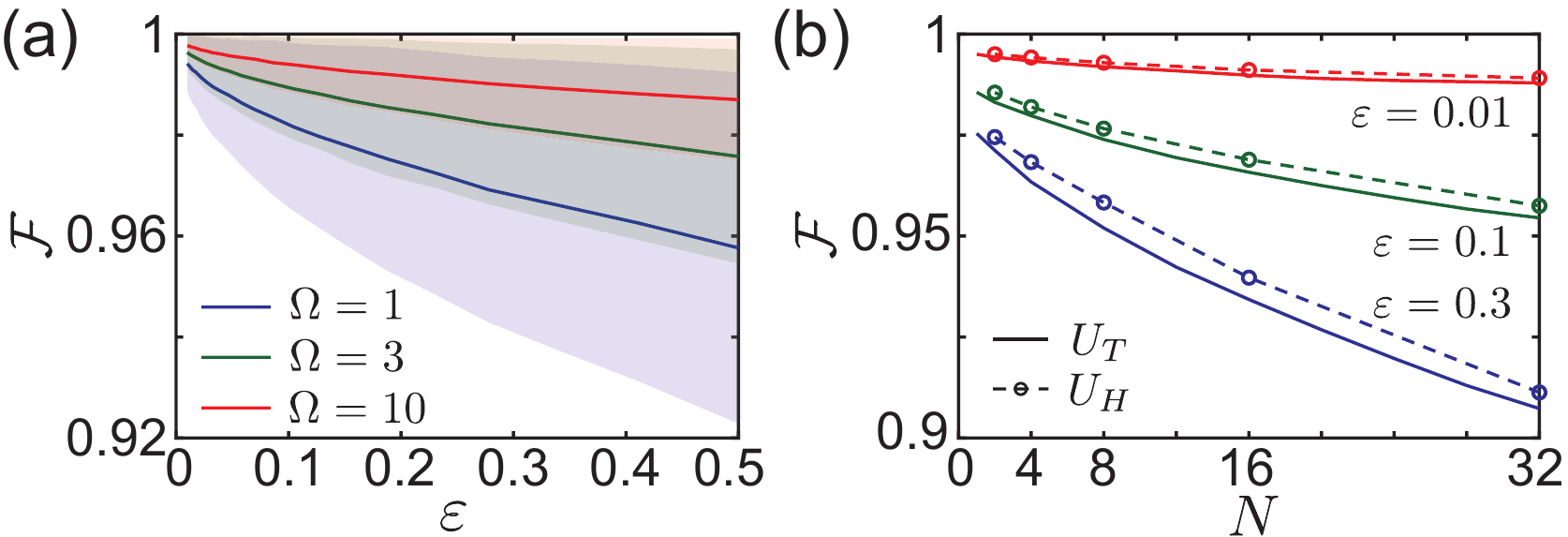}
    \caption{Fidelity of the unitary transformation in the presence of pulse imperfections $\delta g_j(t)$ as defined in Eq.~\eqref{eq:Imperfections}. (a) Scaling of the fidelity for the transformation $U_H(N=4)$ as a function of the strength of the noise $\varepsilon$ and for different bandwidths $\Omega$. 
 The solid lines show the fidelity averaged over $10^4$ noise realizations, with a standard deviation indicated by the shaded areas. (b) Dependence of the fidelity (averaged over $400$ noise realizations) on the number of modes $N$ for different strengths of the noise and for $\Omega=1$. The solid and the dashed lines show the results as obtained for the state-transfer unitary $U_T(N)$ and for the Hadamard transformation $U_H(N)$, respectively.}
    \label{fig4:Imperfections}
\end{figure}

\emph{Imperfections.}---For tunable couplers based on Raman or parametric driving schemes~\cite{hammerer2010,pfaff2016,dassonneville2020,kannan2022}, the complex couplings $g_\mu(t)\sim \Omega_\mu(t)$ are directly proportional to the computer-generated amplitudes of the driving fields, but in practice experimental uncertainties lead to deviations from this exact relation and $g_\mu(t)= g_\mu(t)|_{\rm id}+\delta g_\mu(t)$. To evaluate the impact of such pulse distortions, we assume  
\begin{equation}\label{eq:Imperfections}
    \delta g_{\mu}(t)= \sqrt{\varepsilon\Omega}\int_{-\infty}^{t} ds\, e^{-\Omega (t-s)/2} \xi_\mu(s),
\end{equation}
where the $\xi_\mu(t)$ are independent white noise processes with $\langle \xi_\mu(t)\xi_\nu(t')\rangle = \delta_{\mu\nu} \delta(t-t')$, and $\varepsilon$ and $\Omega$ determine the strength and the frequency range of the pulse distortion.  From the simulations in Fig.~\ref{fig4:Imperfections}(a), we find that the protocol is surprisingly robust with respect to pulse distortions. The infidelity $1-\mathcal{F}$ scales sublinearly with the strength of the noise for rather high values of $\varepsilon$ and fluctuations that are faster than $\gamma_{\rm max}^{-1}$ are further suppressed. Importantly, as shown in Fig.~\ref{fig4:Imperfections}(b), the fidelity of the operation also does not degrade significantly when the number of modes is increased and again a rather weak dependence on $N$ is observed.

In~\cite{supplMultimodeU} we consider also other types of imperfections and show, first of all, that the protocol works equally well for nonidentical modes with frequencies $\omega_\mu=\omega_0+\Delta_\mu$, even for detunings $\Delta_\mu \sim \gamma_{\rm max}$. Therefore, no precise fine-tuning of the local modes is required. Further, the linearity of the transformation makes the protocol insensitive to input noise, such as residual thermal excitations in the channel~\cite{xiang2017,vermersch2017}. The effect of losses, however, reduces the fidelity of the ideal operation $\mathcal{F}_{\rm id}$ to~\cite{supplMultimodeU}
\begin{equation}\label{eq:Fid_losses}
\mathcal{F}\approx  \mathcal{F}_{\rm id} e^{-\gamma t_p} (1-p_{\rm ch} )(1- p_{\circlearrowright})^{2N}.
\end{equation}
Here, $\gamma$ is the bare decay rate of the local mode, $p_{\rm ch}$ is the photon loss probability in the channel connecting register $A$ and register $B$, and $p_{\circlearrowright}$ is the transmission loss of a single circulator or nonreciprocal coupler. While for large $N$, Eq.~\eqref{eq:Fid_losses} places stringent coherence requirements on all network components, it does not reveal any unexpected scalings that depend specifically on the current protocol or on the choice of the unitary transformation. Note, that $\mathcal{F}$ refers to the fidelity of the unitary transformation, as defined in Eq.~\eqref{eq:Fidelity}. The fidelity of the transformed multiphoton state depends in a more complicated way on the initial photon numbers in each mode (see, for example, Ref.~\cite{xiang2017} for the case $N=1$).

\emph{Conclusions.}---In summary, we have presented a universal protocol for implementing excitation-conserving linear unitary transformations between $N$ bosonic modes, where, instead of sending photons through a fixed network of beam splitters and phase shifters, the transformation is implemented through a multiphoton emission and reabsorption process. Therefore, arbitrary unitaries can be realized by simply changing the control pulses and without changing the network configuration. The protocol is robust with respect to the main sources of imperfections and even for very complex unitaries the protocol time only increases linearly with the number of modes.

While the protocol can be implemented with various physical platforms, we envision important near-term applications in the context of circuit QED, where high-fidelity directional couplers and other network components are currently developed~\cite{supplMultimodeU}. Here, the protocol can be used for parallel state-transfer and entanglement distribution schemes or, when combined with local nonlinearities, for large-scale quantum computing architectures based on continuous-variable-encoded qubits.


\begin{acknowledgments}
This work was supported by the European Union's Horizon 2020 research and innovation program under Grant Agreement No. 899354 (SuperQuLAN), the National Natural Science Foundation of China (Grant No. 11874432), the National Key R\&D Program of China (Grant No. 2019YFA0308200), Proyecto Sinérgico CAM 2020 Y2020/TCS-6545 (NanoQuCo-CM) and the CSIC Interdisciplinary Thematic Platform (PTI+) on Quantum Technologies (PTI-QTEP+).
\end{acknowledgments}

\bibliographystyle{apsrev4-2}

\cleardoublepage

\onecolumngrid
\clearpage
\setcounter{equation}{0}
\setcounter{figure}{0}
\setcounter{table}{0}
\makeatletter
\renewcommand{\theequation}{S\arabic{equation}}
\renewcommand{\thefigure}{S\arabic{figure}}
\renewcommand{\bibnumfmt}[1]{[S#1]}
\renewcommand{\citenumfont}[1]{S#1}

\begin{center}
    \textbf{\large Supplementary material for: \\ Universal time-dependent control scheme for realizing arbitrary linear bosonic transformations}
\end{center}

\twocolumngrid

\section{Derivation of the Quantum Langevin Equations}\label{app:QLE}
We consider the network shown in Fig.~1 (a) in the main text, which consists of a set of $2N$ bosonic modes that are weakly coupled to a unidirectional photonic waveguide. In a frame rotating with the bare oscillator frequency $\omega_0$, the coupling between the local oscillators and the waveguide is described by the Hamiltonian
\begin{equation} \label{eq:Interaction Hamiltonian}
    H_{\rm int}(t) = i\hbar\sum_{\mu=1}^{2N}
    \left[ \tilde g_\mu^* (t)    c_\mu  f^\dagger(x_\mu,t)     - \tilde g_\mu (t)   c^\dag_\mu    f(x_\mu,t)   \right],
\end{equation}
where the $x_\mu$ denote the positions of the resonators along the waveguide with $x_\mu> x_\nu$ for $\mu>\nu$.
In Eq.~\eqref{eq:Interaction Hamiltonian}, the bosonic field operator $f(x,t)$ represents the photons in the waveguide.  For a unidirectional channel with a group velocity $v$, this field operator is given by
\begin{equation}
    f(x,t)\simeq   \frac{1}{\sqrt{2\pi}} \int_{\omega_0-B}^{\omega_0+B} d\omega   \, e^{i\omega_0 t} e^{-i\omega(t-x/v)} b_{\omega}(t),
\end{equation}
where $B$ is the bandwidth of the channel and the $b_\omega(t)$ are slowly varying bosonic operators obeying $[b_\omega, b^\dag_{\omega'}]=\delta(\omega-\omega')$. Note that with these conventions the couplings $\tilde g_j$ have dimensions of $\sqrt{{\rm Hz}}$.

The equations of motion for the Heisenberg operators derived from $H_{\rm int}(t)$ are
\begin{eqnarray}
    \dot c_\mu (t)  &=&- \tilde g_\mu(t) f(x_\mu,t),\\
    \dot b_\omega (t) &=& \frac{1}{\sqrt{2\pi}} \sum_\mu  \tilde g_\mu^*(t) c_\mu(t) e^{-i\omega_0 t} e^{i\omega (t-x_\mu/v)}.
\end{eqnarray}
The formal solution of the field operator is
\begin{equation}
    \begin{split}
        &f(x,t)= \tilde f_{0}(x,t) \\
        &+  \sum_\mu  \int_0^t ds\,  \tilde  g_\mu^*(s) c_\mu(s)  \delta_B  \left(t-s-\frac{x-x_\mu}{v}\right)e^{i\omega_0(x-x_\mu)/v}.
    \end{split}
\end{equation}
Here $\tilde f_0(x,t)$ is the free field operator and we have used
\begin{equation}
    \int_{\omega_0-B}^{\omega_0+B} d\omega \,  e^{-i(\omega-\omega_0) (t-s-\tau)}= 2\pi \delta_B (t-s-\tau),
\end{equation}
where $\delta_B(t)$ is the $\delta$-function on timescales that are long compared to the inverse of the channel bandwidth. By reinserting the result for $f(x,t)$ into the equations of motion for the $c_\mu$, we finally obtain
\begin{equation}
    \begin{split}
        \dot c_\mu&= - \tilde g_\mu(t)  \tilde  f_{0}(x_\mu,t) - \frac{|\tilde g_\mu(t) |^2}{2} c_\mu(t) \\
        &-\sum_{\nu<\mu} \tilde g_\mu(t)  \tilde g_\nu^*(t-\tau_{\mu,\nu}) c_k(t-\tau_{\mu,\nu}) e^{i\omega_0 \tau_{\mu,\nu}},
    \end{split}
\end{equation}
where $\tau_{\mu,\nu}=(x_\mu-x_\nu)/v$.

In a final step we switch to rescaled units of time, $t\gamma_{\rm max}\rightarrow t$, where $\gamma_{\rm max}$ is the maximal decay rate, and define the dimensionless couplings $g_\mu(t)=\tilde g_\mu(t)/\sqrt{\gamma_{\rm max}}$. We also define $f_{{\rm in}}(t)= \tilde f_0(x_1,t)/\sqrt{\gamma_{\rm max}}$.
As a result, we obtain the quantum Langevin equations from the main text,
\begin{equation}
    \begin{split}
        \dot c_\mu (t)&= - \frac{|g_\mu(t)|^2}{2} c_\mu(t) -  g_\mu(t) f_{{\rm in},\mu}(t),
    \end{split}
\end{equation}
together with $f_{\rm in,1}(t)\equiv f_{\rm in}(t)$ and $f_{{\rm out},\mu}(t)= f_{{\rm in},\mu}(t)+ g^*_\mu(t) c_\mu (t)$. Note, however, that the relation between incoming and outgoing fields,
\begin{equation}
    f_{{\rm in},\mu}(t)= f_{{\rm out},\mu-1}(t-\tau_{\mu,\mu-1}) e^{i\omega_0\tau_{\mu,\mu-1}},
\end{equation}
still contains propagation delays, which make the set of differential equations nonlocal in time. Because of the unidirectional nature of the channel, these delays can be eliminated by setting $x_1=0$ and defining the time-advanced operators and couplings,
\begin{eqnarray}
    \bar c_\mu(t)&=& e^{-i\omega_0 x_\mu/v} c_\mu(t+x_\mu/v),\\
    \bar g_\mu(t)&=& g_\mu(t+x_\mu/v).
\end{eqnarray}
As a result we obtain the set of equations given in Eq.~(2) and Eq.~(3) in the main text, where we omitted the bar on top of the operators and the couplings for simplicity.


\section{Control pulses} \label{SMSec:ControlPulses}
In this section we present a step-by-step derivation of Eq.~(10) in the main text. This equation provides an explicit solution for the control pulses $g_{B,j}(t)$, which implement a given unitary transformation $U$ for a given set of control pulses $g_{A,j}(t)$.

\subsection{Control pulses for emission}
As a first step in the protocol, we must choose a set of control pulses $g_{A,j}(t)$ for the modes in register $A$. These pulses do not have to be of any specific shape, but for the implementation of a generic unitary transformation the following three conditions must be satisfied:
\begin{enumerate}
    \item The pulses $g_{A,j}(t)$ have a sufficiently large pulse area,
          \begin{equation}\label{eq:LongPulses_supp1}
              \int_{t_0}^{t_f} ds \, |g_{A,j} (s)|^2\gg 1.
          \end{equation}

    \item The pulses $g_{A,j}(t)$ are mutually overlapping,
          \begin{equation}\label{eq:LongPulses_supp2}
              \int_{t_0}^{t_f} ds \,  g_{A,j} (s)  g_{A,k} (s)\neq  0.
          \end{equation}

    \item The pulses $g_{A,j}(t)$ are nonidentical.

\end{enumerate}
As already mentioned in the main text, the first condition ensures that all the initial excitations in register $A$ decay up to exponentially small corrections. For example, for the first mode we find
\begin{equation}
    \langle a_1^\dag a_1\rangle(t_f)= \langle a_1^\dag a_1\rangle(t_0) e^{-\int_{t_0}^{t_f} ds \, |g_{A,1} (s)|^2}.
\end{equation}
For the successive modes the dynamics is more complicated, but the unidirectional propagation of the emitted photons ensures that all the initial excitations eventually leave register A, as long as all the couplings $g_{A,j}(t)$ are switched on for a sufficiently long time.

The second condition is necessary to implement unitary transformations where, for example, a superposition of two or more modes in register A is mapped onto a single mode in register B. This is not possible when the emitted photons have no overlap.

The third condition of non-identical initial pulses is not strictly necessary, but it is included for practical reasons. When applying the numerical algorithm for calculating the optimal pulse shapes as described below, we typically find that non-identical pulses lead to shorter overall protocol times. Also when using identical pulses it is more likely to obtain unphysical oscillations due to numerical errors.

\subsection{Generalized dark state conditions}
To construct the control pulses $g_{B,j}(t)$, we proceed as outlined in the main text and assume that the whole network is initially prepared in the single excitation state  $|\psi_\ell\rangle=\Psi_\ell^\dag|{\rm vac}\rangle$. Here $|{\rm vac}\rangle$ is the $2N$-mode vacuum state and
\begin{equation}
    \Psi _\ell = \sum_{k=1}^N U_{\ell k} a_k(t_0).
\end{equation}
This approach has the conceptual advantage that the whole transfer process can be divided into individual processes, where in each step an initial excitation $|\psi_\ell\rangle$ is mapped onto the single-photon state $b_\ell^\dag |{\rm vac}\rangle$ at the end of the protocol. From a physical perspective it is then clear that this specific process must only involve excitations of the first $N$ modes in register $A$ and the first $\ell$ modes of register $B$. Mathematically, this can be expressed as
\begin{equation}\label{eq:Pk}
    P_\ell(t)= \sum_{\mu=1}^{N+\ell} \left| \left[ \mathcal{G}(t,t_0) \mathcal{U}^\dag \right]_{\mu,\ell} \right|^2 =1,
\end{equation}
where
\begin{equation}
    \mathcal{U}=\left(
    \begin{array}{cc}
            U   & 0_N \\
            0_N & 0_N
        \end{array}\right).
\end{equation}
Eq.~\eqref{eq:Pk} follows from the fact that, given the initial state $|\psi_\ell\rangle$, the population of the $\mu$-th mode of the whole network can be expressed in terms of the Green's function as
$
    \langle c_\mu ^\dag c_\mu\rangle(t)= | \left[ \mathcal{G}(t,t_0) \mathcal{U}^\dag \right]_{\mu,\ell}|^2.
$

The constraint on the excitation probability given in Eq.~\eqref{eq:Pk} can also be rewritten in a differential form as a conservation law,
\begin{equation}\label{eq:PjConservation}
    \dot P_\ell(t)=- | \tilde F_{N+\ell,\ell}(t,t_0)|^2= 0.
\end{equation}
Note that compared to the out-field amplitudes defined in the main text, we use here the slightly different convention
\begin{equation}\label{eq:Outfield_supp}
    \begin{split}
        \tilde F_{\mu ,\ell}(t,t_0)=\,&\langle {\rm vac}| f_{{\rm out},\mu}(t)|\psi_\ell\rangle\\
        =\,& \sum_{\nu =1}^{\mu } g^*_{\nu}(t) \left[ \mathcal{G}(t,t_0) \mathcal{U}^\dag\right]_{\nu,\ell},
    \end{split}
\end{equation}
such that the first index can assume any value $\mu=1,\dots, 2N$. The result in Eq.~\eqref{eq:PjConservation} can be verified by evaluating both sides of the equation. This is most conveniently done by making use of the relation (omitting the time variables)
\begin{equation}\label{eq:dotGU}
    \left[\dot{\mathcal{G}} \mathcal{U}^\dag\right]_{\mu,\ell}= -g_\mu\left( \frac{g^*_\mu}{2}  \left[\mathcal{G} \mathcal{U}^\dag\right]_{\mu,\ell}+ \tilde F_{\mu-1,\ell}  \right).
\end{equation}
Therefore, keeping in mind that $\tilde F_{N+\ell,\ell}(t,t_0)= F_{\ell,\ell}(t,t_0)$, the dark state condition stated in Eq.~(9) in the main text can be directly derived from the conservation of the excitation probability within the first $N+\ell$ modes of the network.

\subsection{Sufficiency of the dark state conditions}
While we have argued that it is necessary to obey the set of dark state conditions $\tilde F_{N+\ell,\ell}(t,t_0)=0$ at any time during the protocol, we now demonstrate that this is even a sufficient requirement for implementing the correct unitary operation. To do so we rearrange Eq.~\eqref{eq:Pk} and show that
\begin{equation}\label{eq:Commutation Relation}
    \left| \left[ \mathcal{G}(t_f,t_0) \mathcal{U}^\dag \right]_{N+\ell,\ell}\right|^2= 1- \sum_{\mu=1}^{N+\ell-1}\left| \left[ \mathcal{G}(t_f,t_0) \mathcal{U}^\dag \right]_{\mu,\ell} \right|^2 = 1.
\end{equation}
For $\ell=1$ this result simply follows from the fact that $G_{AA}(t_f,t_0)=0$ for sufficiently long pulses. Given that $|\left[ \mathcal{G}(t_f,t_0) \mathcal{U}^\dag \right]_{N+1,1}|=1$ and $U$ is unitary,  it  follows that $\left[ \mathcal{G}(t_f,t_0) \mathcal{U}^\dag \right]_{N+1,2}=0$, since the norm of each row of $\mathcal{G}$ is bounded, i.e., $\sum_{\nu=1}^{2N} |\mathcal{G}_{\mu,\nu}|^2\leq 1$. This result then implies that Eq.~\eqref{eq:Commutation Relation} holds also for $j=2$, and so on. Therefore, as long as the control pulses $g_{A,j}(t)$ satisfy Eqs.~\eqref{eq:LongPulses_supp1}-\eqref{eq:LongPulses_supp2} and the set of dark state conditions in Eq.~\eqref{eq:PjConservation} is fulfilled during the whole duration of the protocol, $t\in[t_0,t_f]$, and for all $j=1,\dots, N$, we obtain
\begin{equation}
    [G_{BA}(t_f,t_0)U^\dag]_{\ell,\ell}= e^{i\theta_\ell}.
\end{equation}
The remaining phases can be eliminated by simple rotations of the pulses, $g_{B,\ell}(t)\rightarrow e^{-i\theta_\ell}g_{B,\ell}(t)$, which leave the dark state conditions invariant.  After this adjustment we obtain the desired result, $G_{BA}(t_f,t_0)=U$.

\subsection{Implicit and explicit constructions of the control pulses}
The arguments above show that the targeted unitary transfer operation can be implemented by imposing the set of generalized dark state conditions~\eqref{eq:PjConservation} at all times, but they do not provide a way to construct the required pulses $g_{B,j}(t)$ yet. To show how this can be done,  we first use the input-output relation $f_{{\rm out},N+\ell}(t)=f_{{\rm out},N+\ell-1}(t)+g_{B,\ell}^*(t) b_\ell(t)$ to rewrite the dark state condition $\tilde F_{N+\ell,\ell}(t,t_0)=0$ as
\begin{equation}\label{eq:Implicit}
    g_{B,\ell}^{*}(t)   = - \frac{\tilde F_{N+\ell-1,\ell}(t,t_0)}{\left[\mathcal{G}(t,t_0)\mathcal{U}^\dag\right]_{N+\ell,\ell}}.
\end{equation}
This is still an implicit equation for the control pulses, since $\left[\mathcal{G}(t,t_0)\mathcal{U}^\dag\right]_{N+\ell,\ell}$ depends on $g_{B,\ell}(t)$ as well.
To turn Eq.~\eqref{eq:Implicit} into an explicit equation for the $g_{B,\ell}(t)$, we take its time-derivative and make again use of Eq.~\eqref{eq:dotGU} and  Eq.~\eqref{eq:Implicit} to simplify the result. After some manipulations we obtain the differential equation
\begin{equation}
    \frac{d}{dt} g_{B,\ell}(t)  = g_{B,\ell}(t) \left[ \frac{d}{dt} \log \tilde F^*_{N+\ell-1,\ell}(t,t_0) - \frac{1}{2} |g_{B,\ell}(t)|^2 \right],
\end{equation}
with a nontrivial solution
\begin{equation}\label{eq:Explicit_supp}
    g_{B,\ell}(t)  =
    \frac{\tilde F^*_{N+\ell-1,\ell}(t,t_0) }
    {\sqrt{\int_{t_0}^t ds \, | \tilde F_{N+\ell-1,\ell}(s,t_0)|^2}}.
\end{equation}
This is the result given in Eq.~(10) in the main text.


\section{Numerics}
In this section we provide a more detailed description of the numerical methods that have been used to calculate the control pulses $g_{B,j}(t)$ for all the examples discussed in the main text.

\subsection{Control pulses for register A}
As pointed out above, as long as the initial control pulses $g_{A,j}(t)$ satisfy a few basic requirements, their precise shape is not important for the protocol to work. For all our examples in this work we use the pulses
\begin{equation}
    g_{A,j}(t)=\frac{\eta_j}{\sqrt{e^{-(t-t_c)/\tau}+1}},
\end{equation}
where
\begin{equation}
    \eta_j = \sqrt{\frac{1+(N-j)\delta}{1+(N-1)\delta}}.
\end{equation}

\subsection{Explicit method}

The explicit expression for the control pulses given in Eq.~\eqref{eq:Explicit_supp} is in principle enough to calculate all the $g_{B,j}(t)$ through numerical integration. To do so, one first solves the Green's function $G_{AA}(t,t_0)$ (using, for example, a Runge-Kutta method), which only depends on the known control pulses $g_{A,j}(t)$. With $G_{AA}(t,t_0)$ known, we obtain the out-field amplitude $\tilde F_{N,1}(t,t_0)$ from Eq.~\eqref{eq:Outfield_supp} and the control pulse for the first mode in register B,
\begin{equation}
    g_{B,1}(t) = \frac{\tilde F_{N,1}^*(t,t_0)}{\sqrt{\int^t_{t_0}ds\vert \tilde F_{N,1}(s,t_0)\vert^2}}.
\end{equation}
The knowledge of $g_{B,1}(t)$ can then be used to obtain $\tilde F_{N+1,2}(t,t_0)$ and $g_{B,2}(t)$, etc. However, in practice this method is rather slow for large $N$ as it involves many numerical integrations of the Green's function. 

\subsection{Implicit method}
As an alternative approach to calculate the pulses $g_{B,j}(t)$, we can simply make use of the fact that the set of dark state conditions in Eq.~\eqref{eq:PjConservation} must be satisfied at each point in time. This determines the value of the control pulses through the implicit relation in Eq.~\eqref{eq:Implicit}.

In this method, in a first step we evaluate again the known Green's function $G_{AA}(t_n,t_0)$ and the out-field amplitude $\tilde F_{N,\ell}(t_n)$ on a grid of time points $t_n$ with spacing $\Delta t$. The unknown couplings $g_{B,1}(t_{n})$ and the unknown elements $\mathcal{G}_{N+1,\mu}(t_n,t_0)$ of the Green's function can then be obtained via Euler integration, using Eq.~\eqref{eq:Implicit} to determine the value of $g_{B,1}(t_{n+1})$ for the next time step. Once the values for  $g_{B,1}(t_{n})$ are known, the same procedure can be iterated to obtain $g_{B,2}(t_{n})$, etc.

\subsection{Initial conditions}
A remaining issue for both the explicit and the implicit method is that at the beginning of the protocol the control pulses $g_{B,j}(t)$ are undetermined or can become very large. This arises from the fact that at $t=t_0$ there is a finite out-field from register A, but the population of the $b_1$ mode is still vanishingly small. Therefore, the dark state condition can only be satisfied by a correspondingly large (diverging) coupling.

\begin{figure}[t]
    \includegraphics[width=\columnwidth]{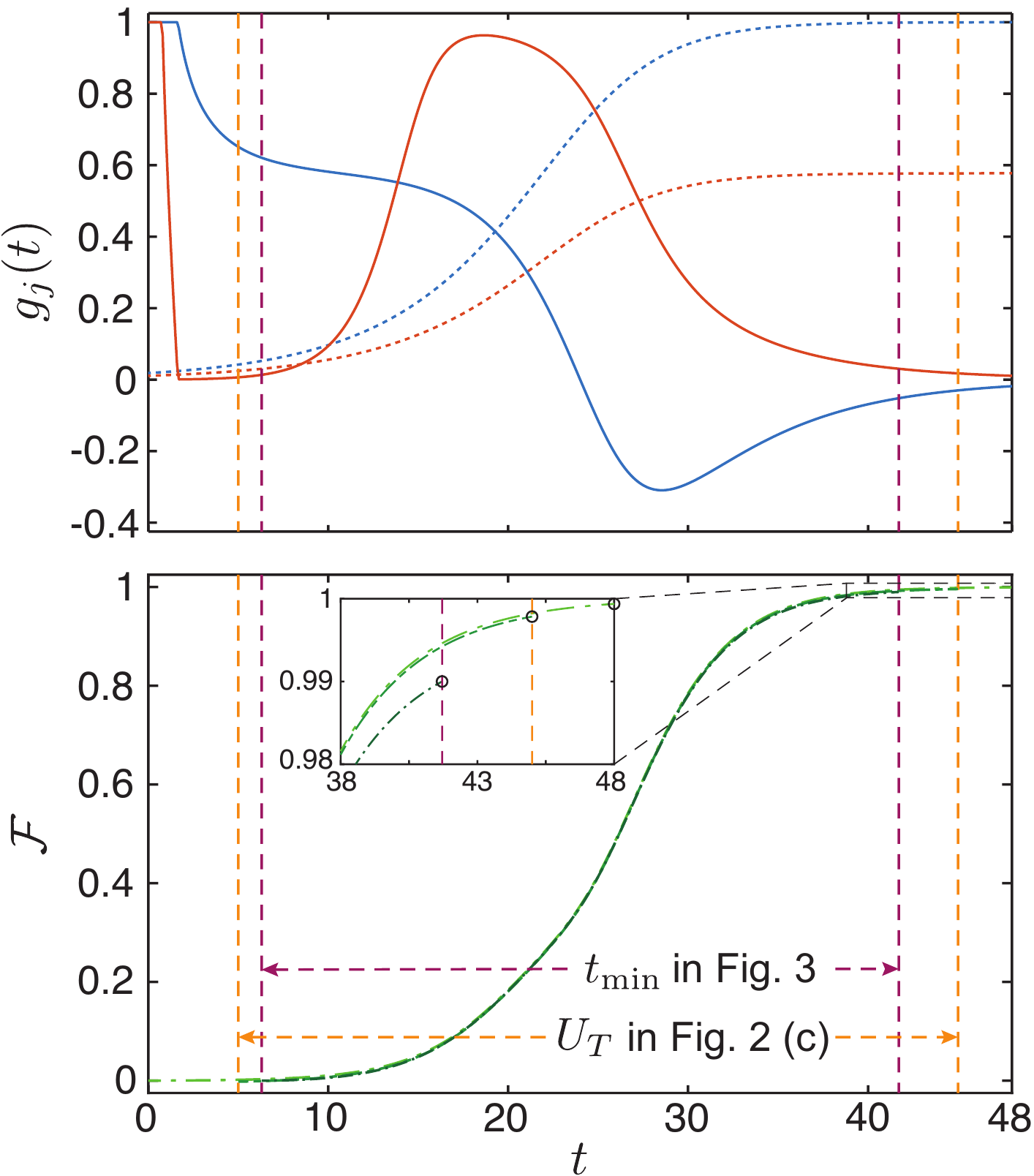}
    \caption{Details of the numerical procedure to obtain control pulses $g_{B,j}(t)$ illustrated for the example $U_T(N=2)$ shown in Fig. 2 (c) in the main text. Upper panel: Starting with the control pulses $g_{A,j}(t)$ (dotted lines) the control pulses $g_{B,j}(t)$ are calculated using either the explicit or implicit method for $t_0'=0$ and $t_f'=48$ and enforcing the bound in Eq.~\eqref{eq:Truncation} at the initial stage. Lower panel: In a second step, keeping the same control pulses, the actual initial time $t_0$ is increased and the final time $t_f$ is decreased (yellow dashed lines) to avoid the initial non-analytic part of the control pulses and set the total protocol time to $t_p=t_f-t_0=40$. To find the minimal time $t_{\rm min}$ plotted in Fig. 3 of the main text, $t_0$ and $t_f$ are further adjusted (purple dashed lines) up to the point where the fidelity drops below a value of $\mathcal{F}=0.99$.}
    \label{fig5:Optimization}
\end{figure}

To deal with this complication in both protocols, whenever $|g_{B,\ell}(t)|>1$ we set the control pulses to a fixed value of
\begin{equation}\label{eq:Truncation}
    g_{B,\ell}(t)\,\,  \rightarrow \,\,  \frac{\tilde F_{N,\ell}^*(t_0,t_0)}{|\tilde F_{N,\ell}(t_0,t_0)|}.
\end{equation}
Although in this case the dark state condition is no longer  fulfilled exactly, this occurs only in the very beginning of the protocol and only causes an exponentially small error for the whole transfer. In the actual numerical simulations, smooth and well-behaved control pulse are obtained as follows: First, by choosing an initial time $t_0'$ and a sufficiently large final time $t_f'$, such that $t_f'-t_0'>t_p$, the control pulses are evaluated according to the prescription in Eq.~\eqref{eq:Truncation}. Then, keeping this set of control pulses fixed, the actual initial time $t_0>t_0'$ and the actual final time $t_f<t_f'$ are chosen such that within the new time window all the pulses are well-behaved, while still reaching the targeted value of the fidelity. This whole procedure is illustrated in Fig.~\ref{fig5:Optimization} for the example shown in Fig. 2 (c) in the main text.

\subsection{Optimized protocol time}
In Fig.~3 in the main text we evaluate the minimal protocol time $t_{\rm min}$ that is required to achieve a fidelity of $\mathcal{F}\geq 0.99$. To do so, we use the implicit method described above and the parameters $\delta=2$, $\tau=1$ and $t_c=t_f/2$ for $t_0'=0$ and an initial protocol time of $t_f'\approx 20 \times [1+(N-1)\delta]$. This results in fidelities of $\mathcal{F}> 0.99$ for all examples. Successively, we run the algorithm for a gradually adjusted $t_0$ and $t_f$, as described above, until the fidelity drops below the threshold (see Fig.~\ref{fig5:Optimization} for an illustrative example).

In the case of the transfer operation $U_T(N)$ we repeat this search for the minimal time $t_{\rm min}$ for different parameters $\delta$. The minimal value of $t_{\rm min}$ obtained in this way, which, for example is reached at $\delta\approx 0.41$ for $N=32$, is marked by the stars in Fig.~3.

\section{Imperfections}\label{sec:Imperfections}
In this section we will discuss in more detail the performance of the protocol in the presence of some of the main imperfections that will be encountered in real experiments. These include:
\begin{itemize} 
\item A static detuning $\Delta_\mu$ of each bosonic mode from the common frequency $\omega_0$.
\item The decay of each boson mode with rate $\gamma$ into a local reservoir different from the waveguide.
\item Propagation losses of photons in the channel connecting register A and register B with probability $p_{\rm ch}$.
\item Photon losses in each circulator or non-reciprocal coupler with probability $p_{\circlearrowright}$.
\end{itemize} 
In addition, static or time-dependent pulse imperfections, $\delta g_{\mu}(t)$, may arise during the conversion of the computer-generated ideal control-pulses into the actual physical couplings. These effects are already discussed in Fig. 4 in the main text. 

\begin{figure}[t]
    \includegraphics[width=\columnwidth]{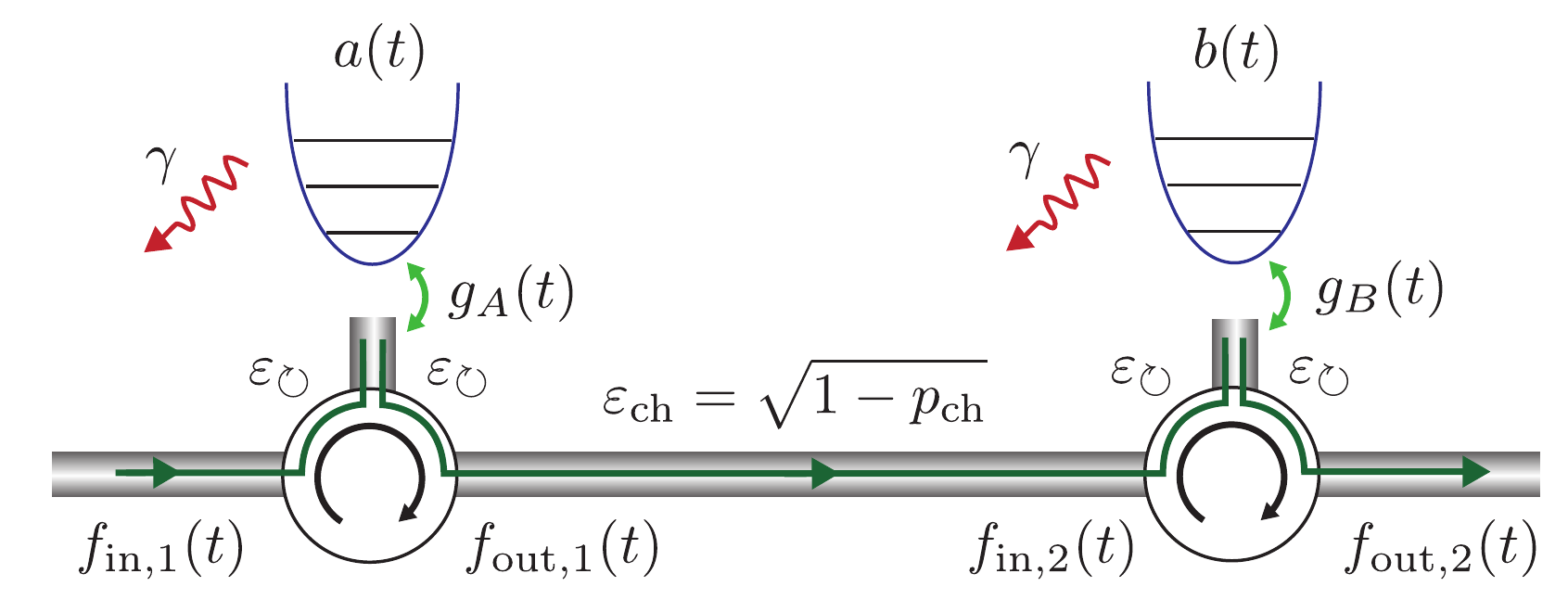}
    \caption{Illustration of the losses considered in the analysis in Sec.~\ref{sec:Imperfections} for the case $N=1$. Every local bosonic mode decays with a rate $\gamma$ and propagation losses between the two registers reduce the field amplitude by $\varepsilon_{\rm ch}=\sqrt{1-p_{\rm ch}}$, i.e. $f_{{\rm in},2}(t)=\varepsilon_{\rm ch} f_{{\rm out},1}(t)$. In addition, whenever fields propagate between two arms of the circulator the amplitudes are reduced by a factor $\varepsilon_\circlearrowright=\sqrt{1-p_\circlearrowright}$.} 
    \label{fig:LossModel}
\end{figure}

To account for the other imperfections, we consider a network model as illustrated in Fig.~\ref{fig:LossModel}. This results in modifications of the quantum Langevin equations and input-output relations given in Eq. (2) and Eq. (3) in the main text and we obtain instead
\begin{equation}
\begin{split}
    \dot{c}_\mu (t)=& \left(-i\Delta_\mu - \frac{\gamma}{2}  -\frac{ |g_\mu(t)|^2}{2}\right)  c_\mu(t)\\
    &- \sqrt{1-p_\circlearrowright} \, g_\mu(t)  f_{{\rm in},\mu}(t)
\end{split}
\end{equation}
and
\begin{equation}
f_{{\rm out},\mu}(t)= (1-p_\circlearrowright) f_{{\rm in},\mu}(t)+ \sqrt{1-p_\circlearrowright} g^*_\mu(t) c_\mu (t).
\end{equation}
In addition, we model the propagation between the two registers by
\begin{equation}
f_{{\rm in},N+1}(t)=\sqrt{1-p_{\rm ch}} f_{{\rm out},N}(t),
\end{equation} 
to account for channel losses. Note that in those modified equations we have omitted all noise operators associated with the additional decay processes. This is valid as long as thermal excitations of these other baths can be neglected. 

When including these modifications, we obtain a new evolution equation for the Green's function, $\partial_t \mathcal{G}(t,t_0)=- \mathcal{M}(t) \mathcal{G}(t,t_0)$, where the matrix elements of $\mathcal{M}(t)$ are given by
\begin{equation}
\begin{split}
&\mathcal{M}_{\mu\nu} (t) =  \left(i\Delta_\mu + \frac{\gamma}{2}\right)\delta_{\mu\nu} \\
&+    g_\mu(t) g^{*}_\nu(t)(1-p_\circlearrowright)^{\mu-\nu} \sqrt{(1-p_{\rm ch})^{\sigma(\mu,\nu)}} \Theta(\mu-\nu).
\end{split}
\end{equation} 
Here, $\sigma(\mu,\nu)=1$ for $\mu>N$ and $\nu \leq N$ and $\sigma(\mu,\nu)=0$ otherwise.

\subsection{Frequency shifts} 
\begin{figure}[t]
    \includegraphics[width=\columnwidth]{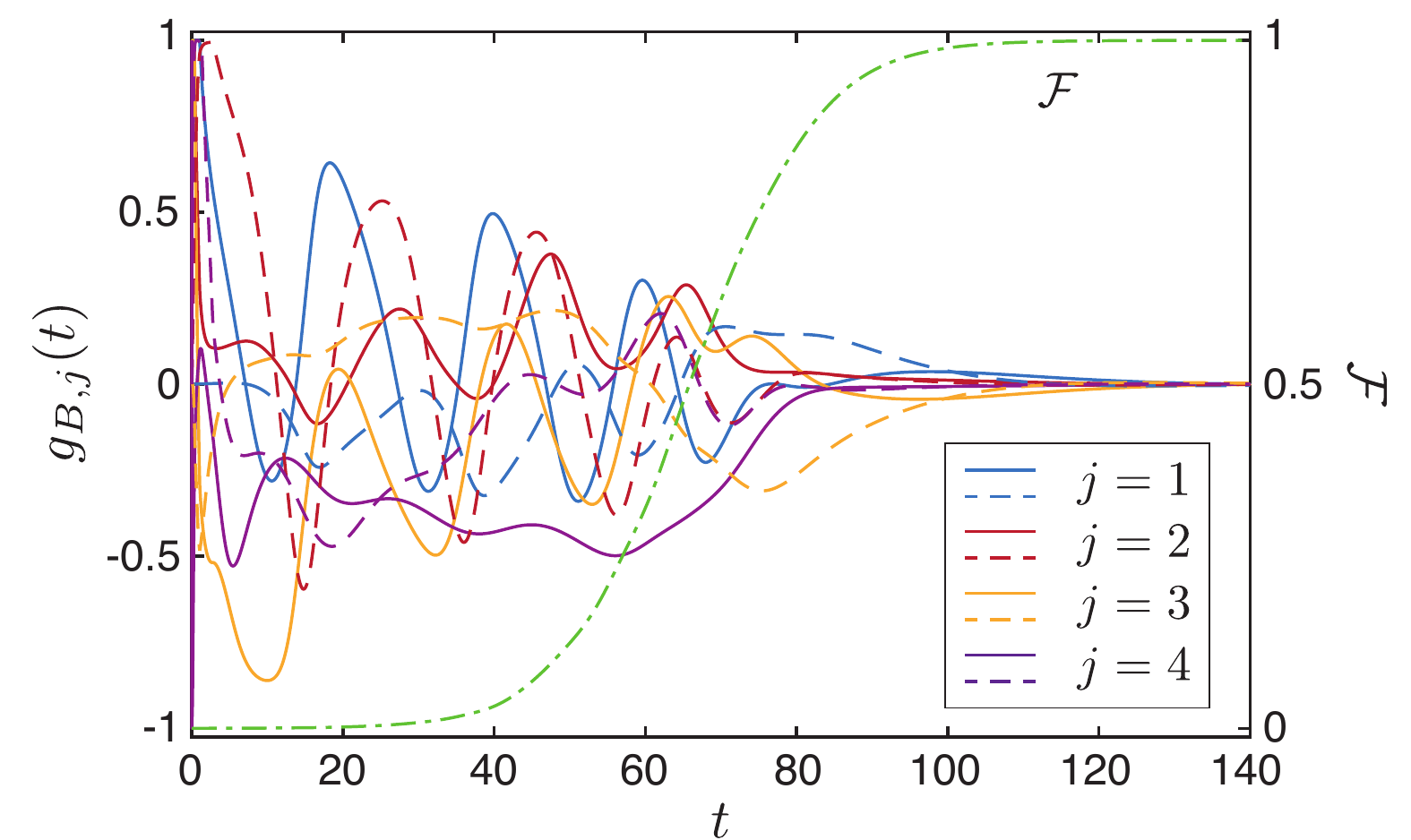}
    \caption{Plot of the control pulses $g_{B,j}(t)$ for implementing the Hadamard unitary $U_H(N=4)$ between four modes. For this plot, the detunings $\Delta_\mu$ of all modes have been randomly chosen from the interval $[-0.5,0.5]$. The solid and the dashed lines show the real and the imaginary parts of the numerically generated pulses $g_{B,j}(t)$, respectively. The control pulse $g_{A,j}(t)$ are the same as in all other numerical examples. The fidelity of the operation is indicated by the dashed-dotted line a reaches a value of $\mathcal{F}\simeq0.999$, which was the targeted accuracy in this simulation.}
    \label{fig:Detunings}
\end{figure}

We first consider the case where losses are negligible, but where each bosonic mode has a different resonance frequency $\omega_\mu= \omega_0+\Delta_\mu$. This means that the relative phases between the modes and also the phases of the emitted photons evolve over time. However, as already shown in Ref.~\cite{xiang2017} for the case of a simple state transfer operation between two modes, i.e., $N=1$, this phase is deterministic and can be compensated by adjusting the phase of the couplings correspondingly. To demonstrate that this strategy works as well for the multi-mode case, we illustrate in Fig.~\ref{fig:Detunings} the implementation of the Hadamard unitary $U_H(N=4)$ for the case where the $\Delta_\mu$ are randomly chosen from an interval $[-0.5, 0.5]$. This plot then shows a set of computer-generated complex control pulses, which implement the operation with the same fidelity as for $\Delta_\mu=0$ and on a similar timescale. We have repeated these simulations for other random unitaries and larger detunings and reached the same performance.

In conclusion, we find that for fixed detunings $\Delta_\mu\sim \gamma_{\rm max}$, the protocol works equally well and there is no need to have exact frequency matching between the modes. In general, for detuned modes, the resulting control pulses will be complex, even when the unitary operation has only real entries. This is not a problem for Raman-type coupling schemes~\cite{hammerer2010,pechal2014} or schemes based on three-wave mixing~\cite{pfaff2016}, where the phase of the coupling is simply given by the phase of the control field. For much larger detunings, $\Delta_\mu \gg \gamma_{\rm max}$, we can move to a rotating frame,
\begin{equation}
c_\mu(t) \rightarrow c_\mu(t) e^{-i\Delta_\mu t},
\end{equation}
and make the ansatz $g_\mu(t) = {\rm g}_\mu(t) e^{-i\Delta_\mu t}$ to remove all fast rotating terms. In this way, we can  control the slowly varying pulses ${\rm g}_\mu(t)$ instead, which simplifies the numerics and the implementation of the control. This procedure works as long as $|\Delta_\mu|\ll B$, which is assumed in the derivation of the quantum Langevin equations in Sec.~\ref{app:QLE}. Therefore, the protocol can also be implemented with far-detuned modes to avoid frequency crowding and unwanted cross-talks between the modes in the same register.

\subsection{Local decay and channel losses} 
Under the assumption of similar local loss rates and a waveguide loss that is relevant only between the two registers, the combined effect of these two dissipation sources is given by
\begin{equation}
G_{BA}(t)= e^{-\frac{\gamma t}{2}} \sqrt{(1-p_{\rm ch})} \left.G_{BA}(t)\right|_{\rm ideal}.
\end{equation} 
Assuming that $\left.G_{BA}(t_p)\right|_{\rm ideal}=U$, the resulting loss-induced reduction of the fidelity $\mathcal{F}$, as defined in Eq. (13) in the main text, is
\begin{equation}\label{eq:Fchannel}
\mathcal{F}= e^{-\gamma t_p} (1-p_{\rm ch}).
\end{equation}
Although this results still depends on the number of modes via the total protocol time, $t_p\sim N$, there are no additional, protocol-specific decoherence effects.

\subsection{Circulator losses} 

\begin{figure}[t]
    \includegraphics[width=\columnwidth]{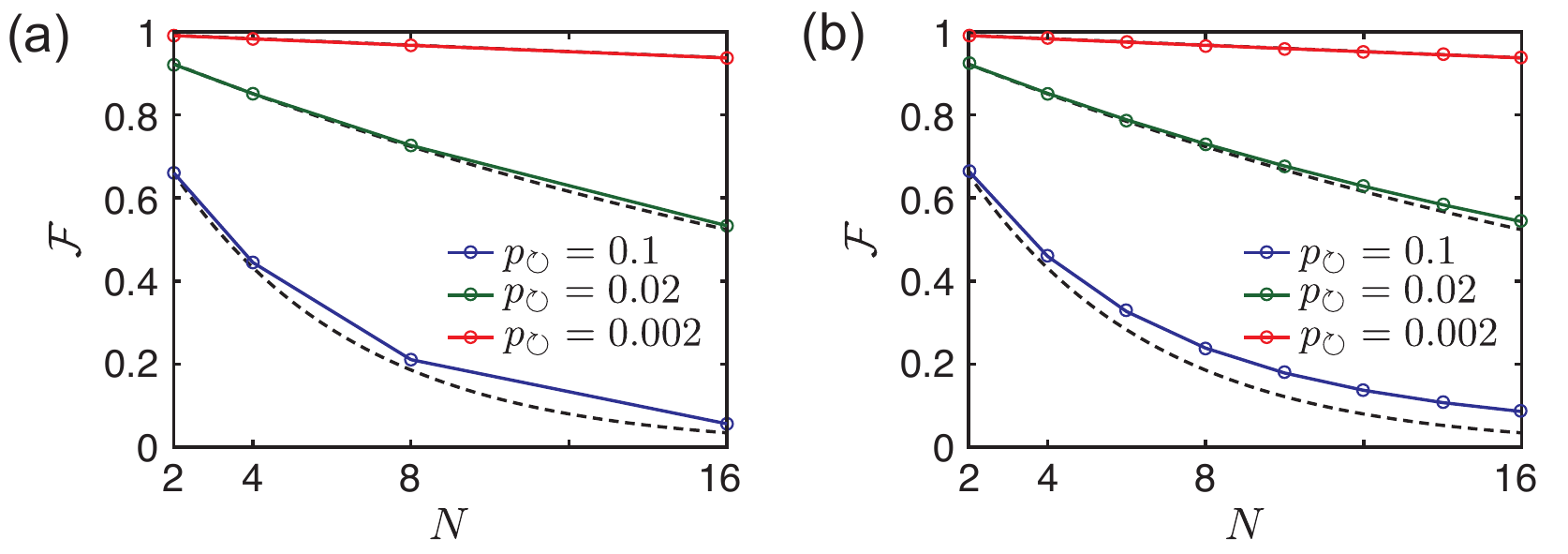}
    \caption{Plot of the fidelity of (a) the Hadamard unitary $U_H(N)$ and (b) the swap operation $U_S(N)$, when taking circulator losses with varying $p_\circlearrowright$ into account. For these plots, a maximal fidelity of $\mathcal{F}|_{p_\circlearrowright=0}=0.999$ and otherwise ideal conditions have been assumed.  The dashed lines show the approximate analytic results given in Eq.~\eqref{eq:Fcirc}, which for $p_\circlearrowright=0.002$ overlaps with the numerical results.}
    \label{fig:Circulator}
\end{figure}

In a next step we assume $\gamma=p_{\rm ch}=0$ and focus solely on the losses induced by the circulators. In this case the resulting fidelity will in general depend on the choice of the unitary transformation. However, for a simple analytic estimate we can consider the state transfer operation $U=\mathbbm{1}$, in which case each photon passes through the same amount of circulators. Therefore, we obtain 
\begin{equation}
G_{BA}(t)= (1-p_{\circlearrowright})^N \mathbbm{1},
\end{equation}
and a reduced fidelity of
\begin{equation}\label{eq:Fcirc}
\mathcal{F}= (1-p_{\circlearrowright})^{2N}.
\end{equation}
As expected, we see that in this case the fidelity depends explicitly on the number of modes, since each node introduces loss for all the photons that pass through it. To confirm this scaling for more general unitaries, we plot in Fig.~\ref{fig:Circulator} (a) and (b) the scaling of $\mathcal{F}$ as a function of $N$ for the Hadamard unitary $U_H(N)$ and the swap unitary $U_S(N)$, respectively. We find that also for these transformations, Eq.~\eqref{eq:Fcirc} provides a very accurate estimate. Therefore, to capture the combined effect of all the considered dissipation sources, we can simply multiply the expressions in Eq.~\eqref{eq:Fchannel} and Eq.~\eqref{eq:Fcirc}, which is the result given in Eq. (15) in the main text.

\subsection{Discussion}
While the importance of different loss channels depends on the specific implementation, a relevant platform for the proposed protocol are superconducting quantum networks. In this case propagation losses can be negligibly small ($<0.01$ dB/m \cite{kurpiers2017}), but the use of typical commercial ferrite circulators with $p_{\circlearrowright}\approx 5-10\%$ still represents a major limitation for current superconducting quantum networks~\cite{magnard2020}. Note that in our convention $p_{\circlearrowright}$ denotes the loss probability when passing between two neighboring legs of a three-port circulator, which matches the single-pass (insertion) losses usually cited in this context. Therefore, Eq.~\eqref{eq:Fcirc} implies that an implementation of a cascaded quantum network using such circulators would be limited to $N\approx 2-4$ nodes, before the fidelity drops below 50\%. 

This estimate, however, represents a worst-case scenario and there are several aspects to keep in mind. First of all, in order to overcome losses in existing circulators, a considerable effort is currently placed on the development of non-reciprocal microwave devices with much better performance. For example, although not tested at the quantum level yet,  commercial ferrite circulators already  achieve insertion losses as low as 0.1 dB ($\sim 2\%)$ and with further optimization losses below $1\%$ are possible~\cite{wang2021}. In addition, there are several schemes for realizing magnetic-free circulators using, e.g., parametric Josephson amplifiers~\cite{sliwa2015,kerckoff2015,chapman2017,lecocq2017,masuda2019}. Currently, these devices are still at a proof-of-concept level, but by relying on low-loss and non-magnetic components only, their performance can reach similar levels as other superconducting circuit components, i.e., $p_{\circlearrowright}\lesssim 10^{-3}$.

Second, while circulators are a conceptually simple way to model cascaded networks, the actual requirement assumed in our analysis is a non-reciprocal resonator-waveguide coupling. For example, in the optical domain, such chiral couplings can be achieved even without circulators by making use of the polarization properties of confined optical beams~\cite{lodahl2017}. Similar type of interactions can be engineered in the microwave regime using non-local couplings~\cite{guimond2020,gheeraert2020}. Using this approach, a directional emission of a single photon state with a fidelity  $>0.95$ has been demonstrated in Ref.~\cite{kannan2022}. Based on the parameters cited on this work, the residual error in this setup seems to be dominated by local decay and consistent with $p_\circlearrowright < 1\%$.

Finally, let us point out that in the derivation of the optimal pulse shapes we have only used the fact that all the photons emitted from register A propagate toward register B. This can be achieved even without circulators by simply reflecting the photons from the left end of the waveguide. This changes the shape of the outgoing wavepackets $\sim \tilde F_{N,\ell}(t,t_0)$, but does not affect the numerical algorithm otherwise (see, for example, Ref.~\cite{lemonde2018} for a related discussion about single-qubit state-transfer protocols in a bidirectional waveguide). Therefore, one can simply omit the circulators in register A, which will improve the overall fidelity by another factor of $\sim2$. However, to be consistent with our original setting, we do not make this change here. 

\bibliographystyle{apsrev4-2}

\end{document}